\documentclass[preprint,prd,showpacs,groupaddress,preprintnumbers,nofootinbib]{revtex4}
\usepackage{graphicx}
\usepackage{epsfig}
\usepackage{bm}

\usepackage{latexsym,amssymb,amsmath,amsfonts,amssymb,wasysym,float}
\usepackage{color}

\newcommand{\En}{E_n}

\newcommand{\Enu}{E_{\overline{\nu}}}

\newcommand{\Enubar}{\epsilon_{\overline{\nu}}}

\newcommand{\cth}{\cos\overline{\theta}_{\overline{\nu}}}

\newcommand{\Enmin}{E_n^{\rm min}}

\newcommand{\epsmin}{\epsilon_{\overline{\nu}}^{\rm min}}
\newcommand{\edeex}{\epsilon_\gamma^{\rm dxn}}

\newcommand{\egdr} {\epsilon_\gamma^{\rm GDR}}

\newcommand{\eamb} {\epsilon_\gamma}

\newcommand{\beq}[1]{\begin{equation}\label{#1}}
\newcommand{\eeq}{\end{equation}}
\newcommand{\bea}[1]{\begin{eqnarray} \label{#1}}
\newcommand{\eea}{\end{eqnarray}}
\newcommand{\ba}{\begin{array}}
\newcommand{\ea}{\end{array}}

\newcommand{\rf}[1]{(\ref{#1})}
\def\be{\begin{equation}}
\def\ee{\end{equation}}
\def\gs{\mathrel{
   \rlap{\raise 0.511ex \hbox{$>$}}{\lower 0.511ex \hbox{$\sim$}}}}
\def\ls{\mathrel{
   \rlap{\raise 0.511ex \hbox{$<$}}{\lower 0.511ex \hbox{$\sim$}}}}

\newcommand{\eps}{\epsilon}

\newcommand{\postscript}[2]{\setlength{\epsfxsize}{#2\hsize}
   \centerline{\epsfbox{#1}}}

\usepackage[usenames,dvipsnames]{xcolor}
\definecolor{rossoCP3}{cmyk}{0,.88,.77,.40}

\begin{document}

\title{\color{rossoCP3} Neutron $\bm{\beta}$-decay as the origin of  IceCube's PeV
    (anti)neutrinos}
\author{Luis A.~Anchordoqui}
\affiliation{Department of Physics and Astronomy, Lehman
  College, City University of New York, NY 10468, USA
}

\date{November  2014}
\begin{abstract}
  \noindent Motivated by the indications of a possible deficit of muon
  tracks in the first three-year equivalent dataset of IceCube we
  investigate the possibility that the astrophysical (anti)neutrino
  flux (in the PeV energy range) could originate from $\beta$-decay of
  relativistic neutrons. We show that to accommodate IceCube
  observations it is necessary that only about $1\%$ to 10\% of the emitted cosmic
  rays in the energy decade \mbox{$10^{8.5} \alt E_{\rm CR}/{\rm GeV} \alt
  10^{9.5}$,} yielding antineutrinos on Earth \mbox{($10^{5.5} \alt
  E_{\overline \nu}/{\rm GeV} \alt 10^{6.5}$),} are observed. Such a
  strong   suppression can  be explained assuming magnetic shielding of the
  secondary protons which diffuse in extragalactic magnetic fields of
  strength $10 \alt B/{\rm nG} \alt 100$  and coherence length $\alt {\rm Mpc}$.
\end{abstract}


\maketitle

\section{INTRODUCTION}

Very recently, the IceCube Collaboration famously announced the
discovery of extraterrestrial neutrinos, including 3 events with
well-measured energies around $10^{6}$~GeV, but notably no events have been
observed above about $10^{6.4}$~GeV~\cite{Aartsen:2013bka}. At $E_{\overline
  \nu}=10^{6.8}$~GeV, one expects to observe a dramatic increase in the
event rate for $\overline \nu_e$ in ice due to the ``Glashow
resonance'' in which \mbox{${\overline \nu_e} + e^- \rightarrow W^-
  \rightarrow {\rm shower}$} greatly increases the interaction cross
section~\cite{Glashow:1960zz}. Indeed, the detection effective area
for $\overline \nu_e$ at the resonant energy is about 12 times that of
off-resonance ($\nu_e, \nu_\mu,\nu_\tau, \overline \nu_\mu, \overline
\nu_\tau$) events. This implies that the falling power law
of the incident neutrino spectrum ($\propto E_\nu^{-\Gamma}$)  is effectively
cancelled and that resonant $\overline \nu_e$ events could have been
seen~\cite{Anchordoqui:2013qsi}.

Various candidate source models have been proposed to explain the
IceCube energy spectrum~\cite{Anchordoqui:2013dnh}. In these models
neutrinos originate dominantly in the decay of pions, kaons, and
secondary muons produced by (photo)hadronic
interactions. Consequently, the expectation for the relative fluxes of
each neutrino flavor at production in the cosmic sources,
($\alpha_{e,S} : \alpha_{\mu,S} : \alpha_{\tau,S}$), is nearly $(1 : 2
: 0)_S$. After neutrino oscillations decohere over the astronomical
propagation distances the flavor conversion is properly described by
the mean oscillation probability. As a result, the flux of ``pionic''
cosmic neutrinos should arrive at Earth with democratic flavor ratios,
$(\alpha_{e,\oplus} :\alpha_{\mu,\oplus} : \alpha_{\tau,\oplus})
\approx (1 : 1 : 1)_{\oplus}$.  If this were the case, then only 1/6
of the total neutrino flux would be subject to the enhancement at the
Glashow resonance. This relaxes the physical significance of the
apparent cutoff.  The obvious question to ask is whether the flavor
ratio $(1:1:1)_\oplus$ is supported by the data.

The IceCube event topologies have been classified as muon tracks and
showers. The full 988-day sample contains 37 veto-passing events (9
tracks and 28 showers) with deposited energies in the range $10^{4.7}
\alt E_\nu/{\rm GeV} \alt 10^{6.3}$. Taken at face value the $9:28$
track-to shower ratio appears consistent with the canonical
$(1:1:1)_\oplus$. However, this is not the case when the atmospheric
muon and neutrino backgrounds are properly accounted for.  The
expected background from atmospheric muons is $8.4 \pm 4.2$ and that
from atmospheric neutrinos is
$6.6^{+5.9}_{-1.6}$~\cite{Aartsen:2013bka}. Altogether, the background
expectation for tracks is about $12$ events, suggesting that the
cosmic component overwhelmingly produces showers inside the
detector. For an unbroken power law energy spectrum with $\Gamma =2$, a
recent statistical analysis indicates that the $(1:1:1)_\oplus$ ratio
is disfavored at the 92\% C.L.~\cite{Mena:2014sja}.\footnote{See,
  however,~\cite{Aartsen:2014tca}.}  The constraint is
lessened by the softer spectra favored by the most recent IceCube
data~\cite{Aartsen:2014muf}. In particular,  for a spectrum $\propto
E_\nu^{-2.3}$, the $(1:1:1)_\oplus$ flavor ratio is disfavored at 86\%
C.L.~\cite{Mena:2014sja}.

It has been suggested that the possible deficit of muon tracks (as
well as the apparent energy gap between $10^{5.5} \alt E_\nu/{\rm GeV}
\alt 10^{6.0}$~\cite{Aartsen:2013bka}) is due to some non-standard physics which favors
Earthly ratios nearly $(1:0:0)_\oplus$, e.g., neutrino
decay~\cite{Beacom:2002vi}, CPT violation~\cite{Barenboim:2003jm},
pseudo-Dirac neutrinos~\cite{Beacom:2003eu}, enhancement of
neutrino-quark scattering by a leptoquark that couples to the
$\tau$-flavor and light quarks~\cite{Barger:2013pla}, sterile neutrino altered
dispersion relations due to shortcuts in extra
dimensions~\cite{Aeikens:2014yga}, and exotic very-soft interactions
of cosmogenic neutrinos~\cite{Illana:2014bda}. In this note we provide
a more mundane explanation, in which a $(3:1:1)_\oplus$ flux of
antineutrinos originates via neutron
$\beta$-decay~\cite{Anchordoqui:2003vc}. The typical energy for the
$\overline \nu_e$ in the lab is that of the parent neutron times
$Q/m_n \sim 10^{-3}$ (the $Q$-value for $\beta$-decay is $m_n - m_p -
m_e = 0.78~{\rm MeV})$. Therefore, to produce PeV antineutrinos we
require a flux of EeV neutrons.  Herein we show that to accommodate
IceCube observations it is necessary that only about 1\% to 10\% of
the emitted cosmic rays in the energy decade $10^{8.5} \alt E_{\rm
  CR}/{\rm GeV} \alt 10^{9.5}$, yielding antineutrinos on Earth
($10^{5.5} \alt E_{\overline \nu}/{\rm GeV} \alt 10^{6.5}$), are
observed. Such a strong suppression can be explained assuming magnetic
shielding of the secondary protons which diffuse in extragalactic
magnetic fields of strength $10 \alt B/{\rm nG} \alt 100$ and coherence
length $\alt {\rm Mpc}$. Before proceeding, we explore the required
assumptions on parameters characterizing the neutron-emitting-sources
(NES).

\section{MODEL ASSUMPTIONS}

We assume that the production of neutrons and photons by cosmic ray
accelerators is a consequence of photo-disintegration of high-energy
nuclei, followed by immediate photo-emission from the excited daughter
nuclei. By far the largest contribution to the photo-excitation
cross section comes from the Giant Dipole Resonance (GDR) at $\egdr
\sim 10~{\rm MeV} - 30~{\rm MeV}$ in the nuclear rest frame.  The
ambient photon energy required to excite the GDR is therefore $\eamb =
\egdr/\gamma_A$, where $\gamma_A=E_A/m_A$ is the boost factor of the
nucleus (of mass number $A$ and charge $Ze$) in the lab.  The GDR decays by the
statistical emission of a single nucleon, leaving an excited daughter
nucleus.  The probability for emission of two (or more) nucleons is
smaller by an order of magnitude.  The excited daughter nuclei
typically de-excite by emitting one or more photons of energies
$\edeex\sim 1-5$~MeV in the nuclear rest frame.  The lab-frame energy
of the $\gamma$-ray is then $E_\gamma=\gamma_A\,\edeex$. To produce
neutrons in the energy range of interest we require a thermal photon
background in the far infrared, $\epsilon_\gamma \sim 10~{\rm meV}$.

There are two channels other than photo-disintegration that might
contribute to $\gamma$-ray and neutrino production. These are
photo-hadronic ($A-\gamma$) and pure hadronic ($A-p$) interactions. In
both cases, $\gamma$-rays (neutrinos) are produced after $\pi^0$
($\pi^+$ and $\pi^-$) decays;  neutrinos carry on average $\sim 1/16$ of the initial
cosmic ray energy per nucleon. To avoid overproduction of neutrinos in
the EeV energy range we assume that collisions of the relativistic
nuclei with the cold ambient interstellar medium are strongly
suppressed, due to an extremely low gas density. Photo-meson production
has a very high energy threshold, being only relevant for very high
energetic beams or in very hot photon environments. Even in these
extreme cases, the fact that this reaction turns on at so high
energies implies that the photons and neutrinos from decaying pions
are produced at very high energies too. The energy threshold for GDR
excitation is more than one order of magnitude below the threshold for
photopion production, $\epsilon_\gamma^{\pi,{\rm th}}  \sim 150~{\rm
  MeV}$. Therefore, in the energy decade of interest ($10^{8.5}~{\rm
  GeV} \alt E_n = E_A/A \alt 10^{9.5}~{\rm GeV}$) our choice of source
parameters automatically suppresses photo-meson production.

Though all cosmic rays experiments point to a dominance of protons
below the ``ankle'' of the cosmic ray spectrum (that is $E_{\rm CR}
\alt 10^{9.6}~{\rm GeV})$, there is a significant disagreement in
interpretation of depth of shower maximum measurements above this
energy, with HiRes~\cite{Abbasi:2009nf} and TA~\cite{Abbasi:2014sfa}
preferring nearly pure protons and Auger~\cite{Aab:2014kda} preferring
a transition to heavies. To remain consistent with the non-observation
of events at the Glashow resonance, the contribution to the cosmic ray
flux from NES cannot extend beyond $10^{9.6}~{\rm GeV}$. This maximum
energy is not inconsistent with the maximum observed energies if one
assumes ultrahigh energy cosmic rays are heavy nuclei, e.g., for iron
nuclei $E_n^{\rm max} \sim 10^{11}/56~{\rm GeV}$.  In the scenario
envisaged here neutron emission from nuclei photo-disintegration
dominates the spectrum below the ankle. The steeply falling neutron
spectrum is overtaken by the harder proton spectrum above about
$10^{9.5}~{\rm GeV}$, where the escape of charged particles becomes
efficient. These overlapping spectra could then carve the ankle into
the spectrum.  The spectrum above the
ankle exhibits a progressive transition to heavy nuclei, as $E_A/Z$
reaches the proton escape energy.  If, on the other hand, the
observed spectrum is dominated by protons above the ankle, we should
then assume that there are two different types of sources contributing
below and above the ankle~\cite{Stanev:2014mla}. This in turn provides
a simple interpretation of the break in the spectrum; namely, a new
population of sources emerges which dominates the more steeply falling
NES population.

\section{FLUX OF ANTINEUTRINOS AND CONSTRAINTS FROM GAMMA RAYS}

We turn to the calculation.  Compared to cosmic distances, the decay of even the boosted neutron
may be taken as nearly instantaneous. Therefore, the basic formula that relates the neutron
flux at the sources to the antineutrino flux observed at Earth
$(dF_{\overline \nu}/dE_{\overline \nu})$ is~\cite{Anchordoqui:2003vc}:
\begin{widetext}
\begin{equation}
 \frac{dF_{\overline \nu} (E_{\overline \nu}) }{d E_{\overline \nu}} =
 \frac{1}{4\pi\,H_0} \! 
  \int dE_n {\cal Q}_n (E_n)  \! \int_0^Q d\epsilon_{\overline
     \nu} \frac{dP   (\epsilon_{\overline \nu})  }{d\epsilon_{\overline \nu}}
 \!  \int_{-1}^1 \frac{d\cos \overline \theta_{\overline\nu}}{2}  
   \; \delta \!\left[E_{\overline \nu}- \frac{E_n \epsilon_{\overline \nu} (1+\cos 
  \overline \theta_{\overline \nu})}{m_n}\right],
\label{nuflux}
\end{equation}
\end{widetext}
where $E_{\overline \nu}$ and $E_n$ are the antineutrino and neutron
energies in the lab, $\overline \theta_{\overline \nu}$ is the
antineutrino angle with respect to the direction of the neutron
momentum in the neutron rest-frame, and $\epsilon_{\overline \nu}$ is
the antineutrino energy in the neutron rest-frame.  The last three
variables are not observed by a laboratory neutrino-detector, and so
are integrated over.  The observable $\Enu$ is held fixed.  The
delta-function relates the neutrino energy in the lab to the three
integration variables, $\Enu = \gamma_n (\epsilon_{\overline \nu} +
\beta \epsilon_{\overline \nu} \cos \overline \theta_{\overline \nu})
= E_n \epsilon_{\overline \nu} ( 1 + \cos \overline \theta_{\overline
  \nu})/m_n$, where $\gamma_n$ is the Lorentz factor and as usual
$\beta \approx 1$ is the particle's velocity in units of $c$. Here,
${\cal Q}_n(E_n)$ is the neutron emissivity, defined as the mean
number of particles emitted per co-moving volume per unit time per
unit energy as measured at the source. In general, the emissivity may
evolve and so depend on time or redshift, but we will ignore this
here. We sum the sources out to the edge of the universe at distance
$H_0^{-1}$ (note that an $r^2$ in the volume sum is compensated by the
usual $1/r^2$ fall-off of flux per source).  Finally, $dP/d\Enubar$ is
the normalized probability that the decaying neutron produces a
$\overline \nu_e$ with energy $\Enubar$ in the neutron
rest-frame. Note that the maximum $\overline \nu_e$ energy in the
neutron rest frame is very nearly $Q$ and the minimum $\overline \nu_e$ energy
is zero in the massless limit. For the decay of unpolarized neutrons,
there is no angular dependence in $dP/d\Enubar$.

Performing the $\cth$ -integration in \rf{nuflux}
over the delta-function constraint leads to
\begin{equation}
\label{nuflux2}
\frac{d F_{\overline \nu}}{d\Enu}(\Enu)   =  \frac{m_n}{8 \pi H_0}\,
\int_{\Enmin} \frac{d\En}{\En}\, {\cal Q}_n(\En)
\int_{\epsmin}^Q \frac{d\Enubar}{\Enubar}\,
\frac{dP}{d\Enubar}(\Enubar)\,,
\end{equation}
with $\epsmin=\frac{\Enu\,m_n}{2\En},$ and
$\Enmin=\frac{\Enu\,m_n}{2Q}$.  An approximate answer is available if
we take the $\beta$--decay as a $1 \to 2 $ process of $\delta m_N \to
e^- + \overline \nu_e,$ in which the antineutrino is produced
mono-energetically in the rest frame, with $\epsilon_{\overline \nu} =
\epsilon_0 \simeq \delta m_N (1 - m_e^2/ \delta^2 m_N)/2 \simeq
0.55$~MeV, where $\delta m_N \simeq 1.30$~MeV is the neutron-proton
mass difference. Setting the beta-decay neutrino energy $\Enubar$
equal to its mean value $\equiv \eps_0$, we have
$\frac{dP}{d\Enubar}(\Enubar)=\delta(\Enubar-\eps_0)$. In the lab, the
ratio of the maximum $\overline \nu_e$ energy to the neutron energy is
$2 \epsilon_0/m_n \sim 10^{-3},$ and so the boosted $\overline \nu_e$'s
have a spectrum with $E_{\overline\nu} \in (0, 10^{-3} \, E_n).$ When
the delta-function is substituted into (\ref{nuflux2}), we obtain
\begin{equation}
 \frac{dF_{\overline \nu} (E_{\overline \nu})}{d E_{\overline \nu}}     = \frac{m_n }{8\,\pi \,\epsilon_0\, H_0}
\int_{E_n^{\rm min}}^{E_n^{\rm max}}
\frac{dE_n}{E_n} \,\, {\cal Q}_n(E_n) \,\, ,
\label{nDK}
\end{equation}
where $E_n^{\rm min} = {\rm max} \{E_{\rm th}^{\rm GDR}, \frac{m_n
  E_{\overline \nu}}{2 \epsilon_0} \}$, and $E_{\rm th}^{\rm GDR} \sim
10^{8.5}~{\rm GeV}$ is the neutron energy from a photo-disintegrated
nucleus at threshold.\footnote{It is implicit that GDR-superscripted
  variables have an $A$-dependence. This could influence the shape of
  the cosmic ray spectrum around $10^{8.5}~{\rm GeV}$, where a
  spectral feature called the ``second knee'' has been reported.}

Next we must relate ${\cal Q}_n$ to an observable. Establishing a
connection between the secondary flux of protons $dF_{\rm CR}/dE_{\rm
  CR}$ and the neutron emissivity is really simple because the
$\beta$-protons, with energies $10^{8.5} \alt E_{\rm CR}/{\rm GeV} \alt
10^{9.5}$, travel undeterred through the universal radiation
backgrounds permeating the universe. However, it is possible that some
protons are shielded by the intergalactic magnetic
field. This will restrict the number of
contributing sources to the cosmic ray spectrum. Including here energy
red-shifting by $1 + z$, we obtain
\begin{equation}
\frac{dF_{\rm CR} (E_{\rm CR})}{dE_{\rm CR}}  = \frac{f}{H_0}
\int_0^{z_{\rm max}} dz \ {\cal Q}_n (1+z, E_n) \,,
\label{protonS}
\end{equation}
where  $f$ is a suppression factor defined as the ratio of
  the observed flux to the one that would be obtained for a continuous
  source distribution without magnetic shielding.

The two observables in (\ref{nDK}) and (\ref{protonS}),
$\beta$-antineutrino and proton spectra at Earth, are related by the
common source. The relation is made explicit by assuming a functional
form for ${\cal Q}_{n} (E_n)$. If we assume a power law with index
$\Gamma$, as shown in~\cite{Weiler:2004jy} the integrals are easily done. For $10^{5.5} \alt
E_{\overline \nu} /{\rm GeV} \alt 10^{6.5}$, we obtain
\begin{equation}
   \frac{dF_{\overline \nu}  (\Enu) }{d\Enu} \approx \frac{10^3}{f}
   \left(\frac{E_{\rm th}^{\rm GDR}}
    {E_n^{\rm max}}\right)^\Gamma 
    \left[\left(\frac{E^{\rm max}_{\overline \nu}}{E_{\overline \nu}}
    \right)^\Gamma -1 \right] \, \ \frac{dF_{\rm CR} (E_{\rm
      th}^{\rm GDR})}{dE_{\rm
     CR}} \,,
\label{znz}
\end{equation}
where 
\begin{equation}
E_{\overline \nu}^{\rm max} = \frac{2 \epsilon_0}{m_n}\,\,  E_n^{\rm
  max} \sim  10^{6.5}\, \,
\left(\frac{E_n^{\rm max}}{10^{9.5}~{\rm GeV}}\right)~ {\rm GeV}\, .
\end{equation}
On the other hand, for $E_{\overline \nu} \alt
2 \epsilon_0 E_{\rm th}^{\rm GDR}/m_n$, the $\overline
\nu_e$ spectrum is flat
\begin{equation}
 \frac{dF_{\overline \nu} (E_{\overline \nu})}{d E_{\overline \nu}}
 \approx \frac{10^{-3}}{f} \  \frac{dF_{\rm CR} (E_{\rm th}^{\rm GDR})}{dE_{\rm
     CR}}   \,,
\end{equation}
because all the free neutrons have sufficient energy, $E_n \agt
10^{8.5}~{\rm GeV}$, to contribute equally to all the $\overline \nu_e$ energy
bins below $E_{\rm th}^{\rm GDR}$.

Taking $\Gamma \simeq 2$ as a reasonable example (\ref{znz}) yields 
\begin{equation}
\left. \frac{E_{\overline \nu}^2 dF_{\overline \nu} (E_{\overline
      \nu})}{dE_{\overline \nu}} \right|_{10^{5.5}~{\rm GeV}} \approx
 \frac{10^{-3}}{f} \ 
({E_{\rm th}^{\rm GDR}})^2 \frac{dF_{\rm CR} (E_{\rm th}^{\rm GDR})}{dE_{\rm
     CR}} \, .
\label{quecagada}
\end{equation} 
Substituting the 
observational value~\cite{Abbasi:2007sv}, 
\begin{equation}
({E_{\rm th}^{\rm GDR}})^2 \frac{dF_{\rm CR} (E_{\rm th}^{\rm GDR})}{dE_{\rm
     CR}} \approx 9 \times 10^{-7}~{\rm GeV} \ {\rm m}^{-2} \ {\rm  s}^{-1} \ {\rm
   sr}^{-1}, 
\end{equation}
into (\ref{quecagada}) and comparing it with the energy square weighted flux reported by the
IceCube Collaboration, ${\cal O} (10^{-8}~{\rm GeV}\, {\rm m}^{-2}\,\,
{\rm s}^{-1}\,\, {\rm sr}^{-1})$~\cite{Aartsen:2013bka}, we require $f
\sim 0.1$ to accommodate our proposal. 

 The propagation of cosmic ray protons in the
  extragalactic magnetic field would be diffusive if the distance from
  the source(s) to Earth is much larger than the scattering
  length. Depending on the magnetic field strength and diffusion
  length, a significant fraction of the ``emitted'' protons can have
  trajectory lengths comparable to the Hubble radius 
  $H_0^{-1}$. However, if the average separation between the sources
  ($d_s \sim n_s^{-1/3}$) in a uniform distribution is much smaller than the
  characteristic propagation length scales due to diffusion and energy
  loss, the observed cosmic ray flux will be the same as that obtained
  for a continuous distribution of sources in the absence of magnetic
  field effects~\cite{Aloisio:2004jda}.  In other words, even at
  energies for which faraway sources do not contribute, as long as the
  observer lies within the diffusion sphere of the nearby sources the
  spectrum is unchanged.  On the other hand, the  flux of
  protons would be suppressed if {\it (i)} particles are unable to
  reach the Earth from faraway sources and {\it (ii)} particles take a
  much longer time to arrive from the nearby sources than they would
  following rectilinear propagation.  It is therefore important to
  study in detail the suppression effect on the closest sources.

\begin{figure}[tbp]
\postscript{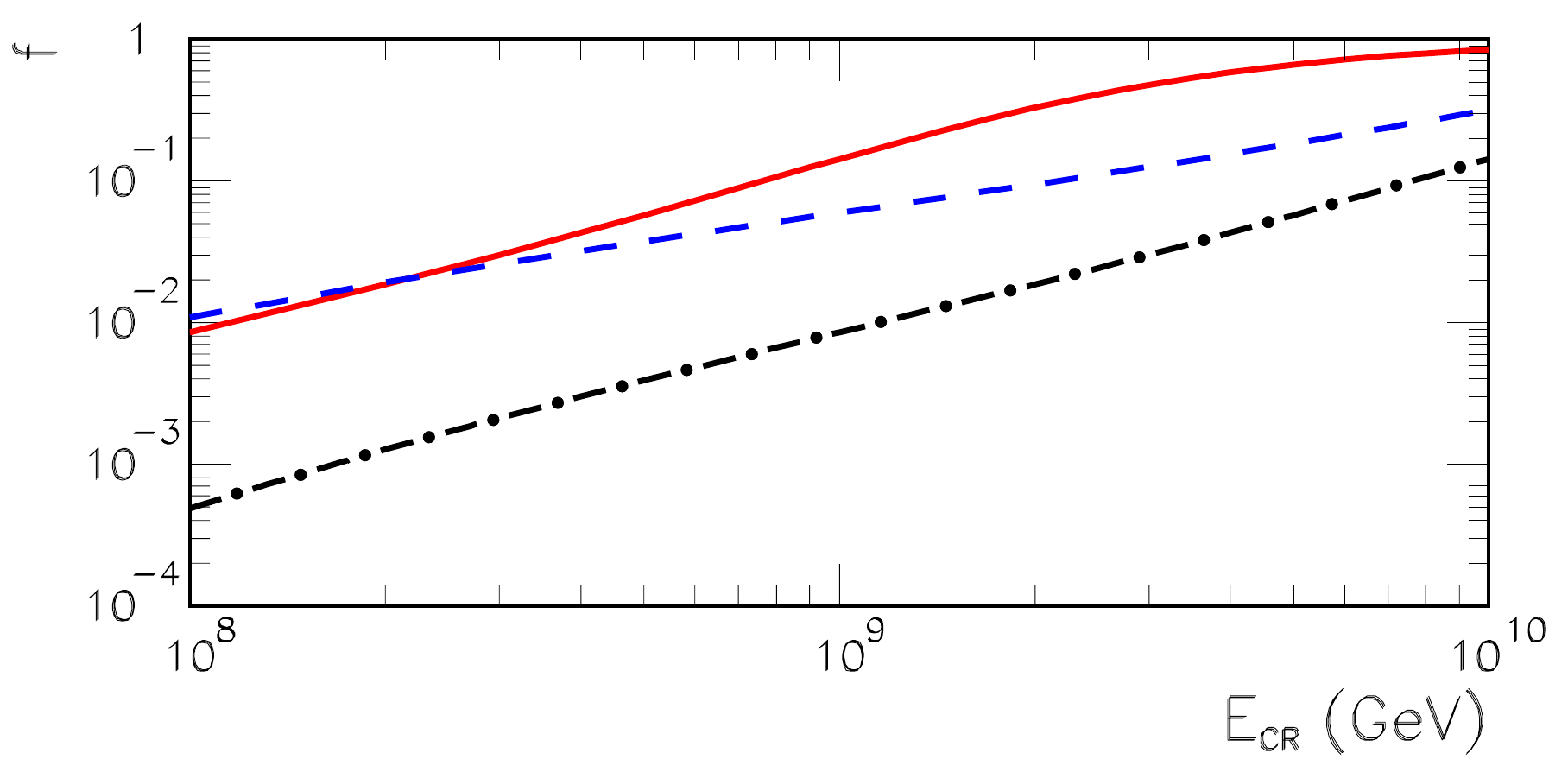}{0.99}
\caption{The suppression factor for various values of the source
  density and magnetic field strength ($n_s/{\rm Mpc}^3,
  B/{\rm nG})$: solid line
 $ (10^{-6}, 10)$, dashed line $(10^{-6.5},100)$, dot-dashed line $(10^{-6},
 100)$. In all cases we have taken $l_c = 1~{\rm Mpc}$.
\label{figura}}
\end{figure}

{}Following~\cite{Lemoine:2004uw} we assume diffusion in a
  random $B$-field with maximum coherent length $l_c$. This assumption
  yields two different propagation regimes depending on the relation
  between the Larmor radius $r_L \simeq 1.1 (E_{\rm CR}/{\rm EeV}) \
  (B/{\rm nG})^{-1}~{\rm Mpc}$ and the coherence length. The
  transition energy between these two regimes, $E_*$, is
determined by the condition \mbox{$r_L(E_*) = l_c$,} yielding $E_\star \simeq
10^9 \ (B/{\rm nG}) (l_c/{\rm Mpc})~{\rm GeV}$. For $\Gamma = 2$,
the suppression factor  can be approximated by
\begin{equation}
f (E_{\rm CR}) \sim {\rm exp} \left[- \left(\frac{a \ d_s}{\sqrt{H_0^{-1}
        l_c}}\right)^\alpha \frac{1}{(E_{\rm CR}/E_*)^\alpha
    + b (E_{\rm CR}/E_*)^\beta} \right] \,,
\label{efe}
\end{equation}
where $\alpha = 1.43$, $\beta = 0.19$, $a =0.2$, $b=0.09$, and
$\sqrt{H_0^{-1} l_c} \simeq 65~{\rm Mpc} \, \sqrt{l_c/{\rm
    Mpc}}$~\cite{Mollerach:2013dza}. In Fig.~\ref{figura} we show three
illustrative examples for which the required range of the value of $f$
can be easily entertained. The approximation in (\ref{efe}) assumes
the magnetic field power to be distributed homogeneously in space. For
inhomogeneous extragalactic magnetic fields, the parameters in
(\ref{efe}) vary significantly depending on the strength of magnetic
fields in the voids of the large scale structure distribution, which
is subject to large uncertainties~\cite{Kotera:2007ca}. We note that
further suppression of the cosmic ray flux can be obtained if some
neutrons decay inside the source, resulting in protons which remain
trapped until attaining the escape energy.

Next, we estimate the $\gamma$-ray flux produced when the
photo-dissociated nuclear fragments de-excite. These $\gamma$-rays
create chains of electromagnetic cascades on the microwave and
infrared backgrounds, resulting in a transfer of the initial energy
into the so-called {\it Fermi}-LAT region, which is bounded by observation~\cite{Ackermann:2014usa}
to not exceed $\omega_{\rm cas} \sim 5.8\times 10^{-7} {\rm
  eV/cm^3}$~\cite{Berezinsky:2010xa}. Fortunately, we can finesse the
details of the calculation by arguing in analogy to the work already done.
The photo-disintegration chain produces one $\beta$-decay antineutrino with
energy of order 0.55~MeV in the nuclear rest frame, for each neutron
produced~\cite{Anchordoqui:2006pd}. Multiplying this result by 2 to include photo-disintegration
to protons in addition to neutrons correctly weights the number of
steps in the chain.  Each step produces on average one photon with
energy $\sim 3$~MeV in the nuclear rest frame.  In comparison, about
12 times more energy is deposited into photons. Including the factor
of 12 relating $\omega_\gamma$ to $\omega_{\bar\nu_e}$, we find from
(\ref{quecagada}) that the photo-disintegration/de-excitation
energy emitted in $\gamma$-rays, $\omega_\gamma \sim 1.1 \times 10^{-7} ~{\rm eV}/{\rm cm}^3$, is below the
{\it Fermi}-LAT bound.\footnote{$\omega_\nu$ is just the area under the
   $E_{\overline \nu}^2\,dF_{\overline \nu}/dE_{\overline \nu}$
  versus $\ln E_{\overline \nu}$ curve~\cite{Berezinsky:2010xa}.}

The analysis described here is subject to several caveats. We have
ignored effects of energy red-shifting of the neutrino and possible
source evolution. A more careful analysis would yield in
(\ref{nuflux}) an additional factor: $H_0 \int dz H^{-1}(z) {\cal Q}_n
(z)/ {\cal Q}_n (0)$.\footnote{A rough estimate can be obtained from
  the following considerations.The redshift from sources at $z=1$ will
  reduce the energy of protons and neutrinos by about 50\% and at
  $z=2$ by about 30\%. If one includes $e^+e^-$ production the energy
  of the protons will be reduced by about 5\% at $z=1$; see Fig.~3
  of~\cite{Unger:2008yz}. Given that protons lose energy during
  propagation scattering off the radiation fields while neutrinos do
  not, the value of $f$ should in fact be somewhat larger than
  computed in the analysis presented here. Additionally calculating
  $f$ precisely requires knowledge of the source evolution.} We have
assumed that not only the nuclei undergoing acceleration remained
magnetically trapped in the source, but also the secondary protons
released in the photo-disintegration process. This may decrease $f$ by
a factor of about 2.  It is also worth stressing that the picture
outlined above is driven by the canonical Fermi index of $\Gamma
\simeq 2$. For $\Gamma = 2.2$, $f$ is reduced by a factor of five and
for $\Gamma = 2.3$, $f$ is reduced by almost one order of magnitude.
Given the current level of uncertainties on the source evolution and
the magnetic horizon, shifting our assumed spectral index from $\Gamma
\simeq 2$ to $\Gamma \simeq 2.3$ will have little impact on the
arguments concerning energetics explored herein. In the future,
improved measurements all-round will require a considerably more
elaborate analysis, including detailed numerical simulations.

It is worth commenting on an additional interesting aspect of this
analysis. Note that $E_{\rm th}^{\rm GDR}$ can be shifted to lower
energies by considering a thermal photon background in the near
infrared, $\epsilon_\gamma \sim 1~{\rm eV}$. Since the cosmic ray
spectrum $\propto E_{\rm CR}^{-3.1}$ is softer than the neutrino
spectrum $\propto E_\nu^{-2.46\pm 0.12}$~\cite{Aartsen:2014muf}, the
source energetics discussed herein would also easily accommodate the
recently proposed {\it two-component} flux model~\cite{Chen:2014gxa},
in which a steeply falling flux of electron antineutrinos populates
the ``low-energy'' range of the cosmic neutrino spectrum observed by
IceCube, and is overtaken at ``high energy'' by a population of
neutrinos produced through pion decay with a harder spectrum.

\section{CONCLUSIONS}

We have presented a model that can accommodate the
apparent deficit of muon tracks in IceCube data without the need of
invoking unknown physics. The model seems unnaturally fine-tuned as it
would be more likely for neutrinos to originate from pion decay; in
particular, the energetics requirement would be more easy to satisfy.
However, Nature is often more subtle than we might like and all options should be considered.
In particular, if the significance of the muon deficit increases as
IceCube collects more data the model presented here will gather plausibility.

\section*{Acknowledgements} I would like to thank Glennys Farrar, Andrei
Gruzinov, Tom Paul, Dmitri Semikoz, and Michael Unger for valuable discussions. I would also
like to thank the Center for Cosmology and Particle Physics at New
York University for their hospitality.  This work was supported in
part by the US NSF (grant CAREER PHY-1053663) and NASA (grant
NNX13AH52G).

\end{document}